\documentstyle[aps,prl,floats,twocolumn,epsf]{revtex}

\def\vrms{v_{\mbox{\tiny{RMS}}}\/}

\begin{document}

\draft

\twocolumn[\hsize\textwidth\columnwidth\hsize\csname@twocolumnfalse%
\endcsname

\title{Screened and Unscreened Phases in Sedimenting Suspensions} \author{Alex
  Levine$^1$, Sriram Ramaswamy$^2$, Erwin Frey$^3$, and Robijn Bruinsma$^4$}

\address{$^1$Exxon Research and Engineering, Rte 22E Clinton Township,
  Annandale NJ 08801, USA,\\ $^2$Department of Physics, Indian
  Institute of Science, Bangalore 560 012, India,\\
  $^3$Institut f\"ur Theoretische Physik, Physik-Department der Technischen
  Universit\"at M\"unchen, D-85747 Garching, Germany,\\
  $^4$Department of Physics, University of California, Los Angeles CA 90095,
  USA.}

\date{\today}
\maketitle
\begin{abstract}
  A coarse-grained stochastic hydrodynamical description of velocity and
  concentration fluctuations in steadily sedimenting suspensions is
  constructed, and analyzed using self-consistent and renormalization group
  methods.  We find that there exists a dynamical, non-equilibrium phase
  transition from a ``unscreened'' phase in which we recover the Calflisch-Luke
  (R.E.\ Calflisch and J.H.C.\ Luke, Phys.\ Fluids {\bf 28}, 759 (1985))
  divergence of the velocity variance to a ``screened'' phase where the
  velocity fluctuations have a finite correlation length growing as
  $\phi^{-1/3}$ where $\phi$ is the particle volume fraction, in agreement with
  Segr\`e {\em et\ al.\/} (Phys.\ Rev.\ Lett. {\bf 79}, 2574 (1997)) and the
  velocity variance is {\em independent\/} of system size.  Detailed
  predictions are made for the correlation function in both phases and at the
  transition.
\end{abstract}

]

Sedimentation \cite{Blanc} is a rich and complex phenomenon in suspension
science and a frontier problem in nonequilibrium statistical mechanics.  The
average sedimentation speed $v_{\rm sed}$ of solute particles drifting down in
a solvent is determined by balancing the driving force (gravity) against the
dissipative force (viscous drag). Giant non-thermal fluctuations in the
velocity and concentration fields in a steadily settling suspension, observed
even for non-Brownian systems, have been a puzzle for some years. Caflisch and
Luke (CL) \cite{Caflisch} showed, for steady sedimentation in a container of
smallest linear dimension $L\/$, that the assumption of purely {\em random\/}
local concentration fluctuations led to velocity fluctuations with a variance
$\langle v^2 \rangle \sim L\/$.  Most experiments, however, find {\em no\/}
dependence of $\langle v^2 \rangle \/$ on $L$ \cite{Nicolai,Xue,Segre},
although Ladd's simulations \cite{Ladd} and the data of Tory {\em et\ al.\/}
\cite{Tory} appear to be consistent with CL.

In this Letter we propose a resolution of this puzzle by means of a set of
coarse-grained, fluctuating nonlinear hydrodynamic equations for the
long-wavelength dynamics of concentration and velocity fluctuations in a
suspension settling steadily in the $-z$ direction, at vanishingly small
Reynolds number.  Our theory is similar in spirit to the Koch-Shaqfeh (KS)
\cite{Koch} ``Debye-like'' screening approach but differs in several important
details and predictions.

The central conclusion of our study is that there are {\em two\/} qualitatively
distinct nonequilibrium phases for a sedimenting suspension. In the
``unscreened'' phase $\langle v^2 \rangle$ diverges as $L$, as in CL and, in
addition, concentration fluctuations with wavevector ${\bf k} = ({\bf
  k}_{\perp}, k_z)$ relax at a rate $\propto \, k^{1/2}\/$. The ``screened''
phase is characterized by a {\em correlation length\/} $\xi$ similar to that
predicted by KS such that $\langle v^2 \rangle \sim L$ for $L \ll \xi$ and
$\langle v^2 \rangle \sim \xi$ for $L \gg \xi$.  Deep in the screened phase we
predict $\xi \sim \phi^{-1/3}$ where $\phi$ is the particle volume fraction.
This is in agreement with the experiments of Segr\`e {\em et\ al.\/}
\cite{Segre}, but not with KS \cite{Koch}. The relaxation rate in the screened
phase is {\em independent\/} of $k$ for $k_z = 0$ and ${\bf k_{\perp}
  \rightarrow 0}$.  Detailed, experimentally testable expressions for the
structure factor and velocity correlations in the screened phase are presented
after we outline our calculations. The two phases are separated in our
``phase-diagram'' (Fig.\ 1) by a striking {\em continuous nonequilibrium phase
  transition\/} where $\xi$ diverges at least as rapidly as $\left( K - K_c
\right)^{-1/3}$ as a control parameter $K$ is decreased towards a critical
value $K_c$.

The hydrodynamic equations we used to arrive at these results are
\begin{equation}
\label{main}
\frac{\partial c}{\partial t} + {\bf v} \cdot \bbox{\nabla} c =
[D_{0\perp} \nabla_{\perp}^2 + D_{0z} \nabla_z^2]c +
\bbox{\nabla} \cdot {\bf f}({\bf r},t)
\end{equation}
and
\begin{equation}
\label{main2}
\eta \nabla^2 v_i({\bf r},t) = m_R g
P_{iz} c({\bf r},t),
\end{equation}
where $c({\bf r}, t\/)$ and ${\bf v}({\bf r}, t )\/$ are the fluctuations about
the mean concentration $c_0$ and the mean sedimentation velocity $- v_{\rm sed}
\hat{z}$ respectively.  We justify these equations briefly below; for a more
detailed discussion we refer the reader to Ref.~\cite{levine}.  Eq.~\ref{main}
is the anisotropic randomly forced advection-diffusion equation with bare
uniaxial diffusivities $( D_{0 z}, D_{0 \perp} )$ and a random stirring force
${\bf f}({\bf r},t)$ \cite{mult}. The Stokes equation, Eq.~\ref{main2}, which
expresses the balance between the driving by gravity and the dissipation by the
viscosity $\eta$, describes how the concentration fluctuations produce velocity
fluctuations.  Here $m_R g\/$ is the buoyancy-reduced weight of a particle,
while the pressure field has been eliminated by imposing incompressibility via
the transverse projection operator $P_{ij} = \delta_{ij} - \nabla_i \nabla_j
(\nabla^2)^{-1}\/$.

Hydrodynamic equations such as Eqs.~\ref{main} and \ref{main2} arise from a
coarse-graining of the microscopic equations of motion. The latter, for the
main case of interest here, {\em viz.\/}, non-Brownian suspensions at zero
Reynolds number, are the deterministic equations of Stokesian dynamics for $N$
hydrodynamically coupled particles, and are known to be chaotic \cite{chaos}.
The noise, or random stirring current ${\bf f}({\bf r},t)$ and the
diffusivities in Eq.~\ref{main} represent a phenomenological description of the
deterministic chaos at length scales below the coarse-graining length $\ell$
(which must be large compared to the particle radius $a$).  We use these
hydrodynamic equations to predict the velocity and concentration fluctuations
at length scales large compared to $\ell$ driven by the random stirring at
short distances.

We assume, as is reasonable, that ${\bf f}({\bf r},t)$ is Gaussian white noise
with uniaxial symmetry:
\begin{equation}
\label{corr}
\langle  f_i({\bf r},t) f_j({\bf r'},t')  \rangle =
2 c_0 N_{0}^{ij} \delta({\bf r}-{\bf r'}) \delta(t-t')
\end{equation}
with an anisotropic noise amplitude $ N_{0}^{ij} =
N_{0\perp}\delta^{\perp}_{ij} + N_{0z} \delta^z_{ij}$, where $\delta^z_{ij}$
and $\delta^{\perp}_{ij}$ are the projectors along and normal to the $z$ axis,
respectively.  Because of the nonequilibrium origin of the noise and diffusion
constants, we may not\cite{driven} assume that $N_{0 \perp}/N_{0 z} = D_{0
  \perp}/D_{0 z}$ as would be true for the Langevin equation of a dilute
suspension at thermal equilibrium. Note that no correlations have been fed in
via the noise: any that emerge in the long-wavelength properties are a result
of the interplay of advection and diffusion.

Let us now consider the nature of the spatio-temporal correlations implied by
Eqs.~\ref{main} and \ref{main2}.  We will focus on the structure factor for
concentration fluctuations
\begin{equation}
\label{sq}
S(q) \equiv c_0^{-1} \int d^dr \langle c({\bf 0}) c({\bf
r})\rangle e^{-i {\bf q}.{\bf r}}
\end{equation}
from which the velocity structure factor can be derived through
Eq.~\ref{main2}.  If we ignore the advective nonlinearity ${\bf v} \cdot
\bbox{\nabla} c$, then $S({\bf q})$ can be computed by straightforward Fourier
transformation of Eq.~\ref{main}, resulting in
\begin{equation}
\label{sq:2}
S({\bf q}) = S_0({\bf q}) \equiv \frac{N_{0 \perp} q_\perp^2 +
N_{0 z} q_z^2}{D_{0
\perp} q_\perp^2 + D_{0 z} q_z^2}.
\end{equation}
Using Eq.~\ref{sq:2} in Eq.~\ref{main2} we can compute $\langle v^2 \rangle$ as
a function of the system size $L$ with the result:
\begin{equation}
\label{vel_var}
\langle v^2 \rangle \sim \int_{q > 1/L} \frac{S({\bf q})}{q^4}
\sim L.
\end{equation}
In other words, neglecting large-scale advection by the velocity fluctuations
leads to the CL \cite{Caflisch} result.

To include the effect of the advective nonlinearity we have performed a
self-consistent mode coupling calculation \cite{gal} on
Eqs.~\ref{main}-\ref{corr}.  Our results can be expressed in terms of a {\em
  renormalized\/} relaxation rate
\begin{equation}
\label{rateren}
R({\bf q}) = D_{\perp}({\bf q}) q_{\perp}^2 + D_z({\bf q})
q_z^2 +  \Gamma({\bf q})
\end{equation}
and a {\em renormalized\/} structure factor of the form
\begin{equation}
\label{sq:new}
S({\bf q}) = \frac{N_\perp({\bf q}) q_\perp^2 + N_z({\bf q})
q_z^2}{R({\bf q})}.
\end{equation}
The quantities $D_{z,\perp}({\bf q})$ and $N_{z,\perp}({\bf q})$ represent
renormalized diffusivities and noise amplitudes \cite{freq}.  But, most
importantly, the advective nonlinearity to lowest-order perturbation theory
leads to an additional term in the renormalization of the relaxation rate which
is of the form $\Gamma({\bf q}) = \gamma({\bf q}) q_{\perp}^2/q^2$.  Starting
from the stochastic hydrodynamic equations, Eqs.\ref{main}-\ref{corr}, it turns
out that the amplitude of this singular contribution becomes a constant,
$\lim_{q \rightarrow 0} \gamma ({\bf q}) \propto I(\beta_N, \beta_D)$, which
depends on the anisotropy ratios of the noise and diffusivity coefficients
\begin{eqnarray}
\label{beta}
\beta_N = \frac{N_\perp}{N_z}, \quad {\rm and} \quad 
\beta_D = \frac{D_\perp}{D_z}.
\end{eqnarray}
In particular $I(\beta_N, \beta_D)$ is proportional to $\beta_N - \beta_D$.
and consequently may change sign upon varying the noise and diffusivity ratios.
For $I(\beta_N, \beta_D) < 0$ this would lead to exponentially growing
concentration fluctuations in the limit of long wavelength. Here we do not
pursue this intriguing possibility further but instead restrict our attention
to $I(\beta_N, \beta_D) \geq 0$, for which the model can either be treated
within dynamic renormalization group theory or using self-consistency methods.

We start our discussion at the borderline of stability, $\beta_N = \beta_D$.
For these parameter values it can be shown that the fluctuating hydrodynamic
equations describe a dynamics which obeys detailed balance
\cite{detailed_balance}: the advective nonlinearity does not affect the
equal-time correlations, and $S({\bf q})$ in particular is just the constant
$N_{\perp}/D_{\perp}$.  There are singularities in $N_{\perp,z}$ and $D_{\perp,
  z}$ which we discuss later.

For $\beta_N \geq \beta_D$, detailed balance is violated and a singular
diffusion term $\Gamma ({\bf q})$ is generated within perturbation theory. In
order to analyze the dynamics in this regime we use one-loop self-consistent
theory (mode coupling theory) and arrive at the expression
\begin{eqnarray}
\label{gamma:eq}
\Gamma({\bf q}) = && c_0 \left(\frac{m_R g}{\eta}\right)^2 
\int_k \frac{q_i P_{iz}({\bf
k})k_j P_{jz}({\bf q})}{k^2 q^2} \nonumber \\ \times
&&\frac{\left[S ({\bf q - k}) - S ({\bf  k})\right]}{R({\bf
k}) + R({\bf q - k})}
\end{eqnarray}
with $R({\bf q})$ given by (\ref{rateren}), and similar self-consistent
integral equations for $D_\perp({\bf q})$, $D_z(\bf q)$, $N_\perp(\bf q)$, and
$N_z(\bf q)$. We find that there are two types of iteratively stable solutions
to these coupled self-consistent equations: those with $\gamma(q \rightarrow 0)
> 0$, which we obtain below the solid line in the phase diagram spanned by the
two anisotropy ratios (``screened'' phase in Fig.\ 1), and those with $\gamma(q
\rightarrow 0) = 0$, which arises for values of the anisotropy parameters that
lie above the solid line and below the dashed line of the same figure, i.e., in
the ``unscreened'' phase. Note that within the self-consistent theory the line
in the phase diagram where $\gamma(q=0)$ changes sign (solid line) has shifted
with respect to the result of the one-loop perturbation theory discussed above
(dashed line).

\paragraph*{Screened Phase:}
In the screened phase, $\Gamma({\bf q})$ is of the form $\gamma q_\perp^2/q^2$
in the small $q$ limit, with $\gamma$ a finite constant. This implies that the
structure factor at small wavenumber becomes
\begin{equation}
\label{structure:gamma}
S({\bf q}) \simeq \frac{N_\perp q_\perp^2 + N_z q_z^2}{D_\perp q_\perp^2
+ D_z q_z^2 + \gamma q_\perp^2 / q^2}
\end{equation}
with $N_{\perp,z}$ and $D_{\perp,z}$ constants. From Eq.~\ref{structure:gamma}
we can define a correlation length $\xi \equiv \left(
  D_\perp/\gamma\right)^{1/2}$ such for $q_\perp \gg 1/\xi$ the structure
factor is not significantly affected by advection.  On the other hand, for $
q_\perp \ll 1/\xi$ the in-plane structure factor reads $S({{\bf q}}_\perp, q_z
= 0) \simeq \left(N_\perp/\gamma \right) q_\perp^2$, while $S({{\bf q}}_\perp =
0, q_z) \simeq \left(N_z/D_z \right)$.  Physically, this means that at long
wavelength advection strongly suppresses in-plane concentration fluctuations.

Using Eq.~\ref{structure:gamma} in conjunction with Eq.~\ref{main2}, one finds
that for length scales $L$ less than $\xi$, $\langle v^2 \rangle \propto L$,
consistent with CL, while for $L$ large compared to $\xi$, $\langle v^2 \rangle
\propto \xi$.  Velocity fluctuations on length scales small compared to $\xi$
are thus highly correlated while they become uncorrelated at larger length
scales.

Deep inside the screened phase, i.e., for large $\gamma$, the renormalization
of the diffusion and noise parameters is negligible and we can explicitly
compute $\gamma$, and thus $\xi$, by inserting Eq.~\ref{sq:new} in
Eq.~\ref{gamma:eq} using the bare values for the $N$'s and $D$'s.  We find
\begin{equation}
\label{eq:xi}
\xi = 8 (\frac{m_R g}{ \eta D})^{-2/3} c_0^{-1/3}  \left( 1  -
\frac{2}{\beta_N} \right)^{-1/3},
\end{equation}
where for simplicity we have set $D_{0 \perp} = D_{0 z} = D$.  According to
Eq.~\ref{eq:xi}, the correlation length increases as we decrease the $\beta_N$
parameter (which could be done by increasing the {\em thermal\/} noise
amplitude) and diverges at $\beta_N = 2$.  Strictly speaking, as $\beta_N
\rightarrow 2$, the diffusivity corrections are no longer negligible, and the
actual divergence of $\xi$ is probably stronger than (\ref{eq:xi}), and occurs
at a larger value of $\beta_N$.  An explicit analytical (but lengthy) result
for the correlation length $\xi$ can also be obtained throughout the screened
phase as a function of both anisotropy parameters\cite{levine} and the phase
boundary can also be computed.  The phase boundary resulting from this result
is shown in Fig.\ 1 as the solid line separating the screened from the
unscreened phases.  The dashed line in the figure corresponds to the set of
parameter values where the hydrodynamic equations correspond to a Langevin
dynamics in thermal equilibrium.

\paragraph*{Unscreened Phase:} As already noted above, the hydrodynamic
equations obey detailed balance \cite{detailed_balance} along the line $\beta_N
= \beta_D$ in the phase diagram.  As a consequence the ratio of noise to
diffusivity can be identified as a direction-independent ``noise-temperature''.
Furthermore, the structure factor $S (\bf q)$ becomes a constant
$D_\perp/N_\perp$ and we recover the CL result.  In conjunction with an
exponent identity resulting from Galilean invariance this is enough to
determine the dynamic exponent exactly, $z=d/2-1$. This implies that the
diffusivities and noise amplitudes scale as $q^{-\epsilon/2} = q^{-3/2}$ for
long wavelength. Even though there are now singular corrections to
$D_{z,\perp}(\bf q)$ and $N_{z,\perp}(\bf q)$, the anomalous $\Gamma(\bf q)$
term is zero.  For parameter values in the regime between the dashed line
(detailed balance line) and the solid line, which marks the location of the
nonequilibrium phase transition, renormalization group methods may be used to
determine the renormalization of the noise and diffusivity amplitudes. In view
of the results from the above self-consistency calculation ($\gamma =0$ in the
unscreened phase) and the exact results at the detailed balance line it is
quite likely that the resulting renormalization group flow will tend towards a
fixed point which obeys detailed balance. We leave the details of such an
investigation for a future publication \cite{levine}.

\begin{figure}
\epsfxsize=0.95\columnwidth
\centerline{\epsfbox{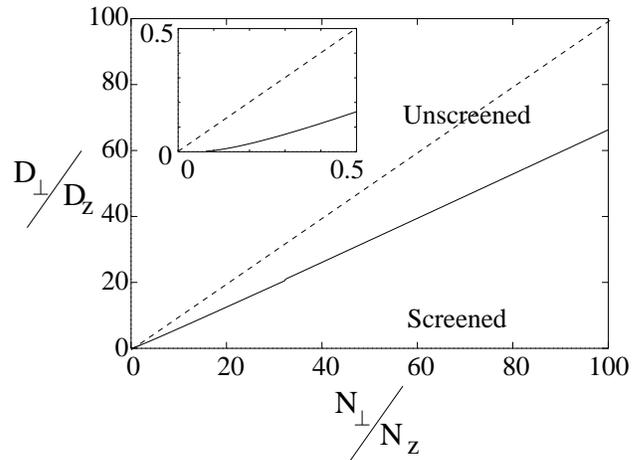}}
\label{phdiag}
\caption[]{Dynamical phase diagram for sedimentation.  Below the solid line
  in the parameter space spanned by the anisotropy factors for the noise and
  diffusivities velocity and concentration fluctuations have a finite screening
  length in the limit vanishing wavevector.  This region is called ``screened''
  above.  In the upper region called ``unscreened'' the screening length
  becomes infinite.  The dashed line represents the set of values for the
  anisotropy factors where the hydrodynamics obeys detailed balance.  The inset
  shows the behavior of the phase boundary in the limit of large noise and
  diffusivity in the vertical direction as compared to the horizontal plane.}
\end{figure}

The analysis of our hydrodynamic equations thus confirms that screening can
suppress the CL divergence of $\langle v^2 \rangle$ with $L$, as argued by KS,
while it allows for a second, unscreened phase.  This result may help explain
the conflicting results on $\langle v^2 \rangle$ obtained by different workers
\cite{Nicolai,Xue,Segre,Ladd,Tory}. The self-consistent structure factor,
Eq.~\ref{sq:2} we obtained differs significantly from the one proposed by KS.
Experimental test will thus be of considerable importance.  Measurements of
$S({\bf q})$, for example by PIV \cite{piv} (Particle Imaging Velocimetry),
would constitute the most direct test of the theory since our prediction that
$S({\bf q}_\perp, q_z = 0) \propto q_\perp^2$ does not hold in the KS
description. Detailed measurements of $S({\bf q})$ for sedimenting solutions
are not yet available.  However, Segr\`e {\em et\ al.\/} \cite{Segre} do report
that the size-dependence of the amplitude $\langle v^2 \rangle$ of the velocity
fluctuations depends on a characteristic length scale $\xi_S$ such that
$\langle v^2 \rangle \propto \xi_S$ for length scales $L \gg \xi_S$ while for
$L \ll \xi_S$, $\langle v^2 \rangle$ grows with $L$.  They report that $\xi_S
\sim a \phi^{-1/3}$ with $\phi$ the particle volume fraction.

Our correlation length $\xi$, in Eq.~\ref{eq:xi}, has the same physical
interpretation as $\xi_S$. Deep in the screened phase, i.e., for $I(\beta_N,
\beta_D ) \gg 0$, $\xi$ can be written as:
\begin{equation}
\label{answer}
\xi(\phi) \sim (m_R g / \eta D)^{-2/3} 
a \phi^{-1/3} I(\beta_N, \beta_D )^{-1/3}
\end{equation}
On scaling grounds, we expect that $D \propto \delta \, \vrms \xi$ with $\delta
\, \vrms$ the root mean square of the velocity field fluctuations.
Experimentally, $\delta \, \vrms \xi$ is found to be independent of volume
fraction $\phi$.  In that case, Eq.~\ref{answer} reproduces the experimentally
observed volume-fraction dependence, in contrast to KS \cite{Koch}.  It should
be noted that this volume fraction dependence of the correlation length implies
that there is a fixed number of colloids within a correlation volume
independent of volume fraction.

The observation of a transition from the screened to the unscreened phase would
obviously be the most conclusive evidence supporting our theory, in particular
if the transition were accompanied by a divergence of the velocity fluctuation
correlation length. Even in the absence of such direct evidence, the
observation of screened behavior combined with our theory requires that the
anisotropies in the noise and diffusivity lie in the lower region of our
dynamical phase diagram, Fig.\ 1.  A complete test of our theory thus requires
measurement of the $N$ and $D$ parameters.  These could be obtained from the
measurement of the steady-state static structure factor $S({\bf q})$, {\em
  e.g.\/} by particle imaging or light scattering experiments both along the
$z$ direction and in the $x-y$ plane, coupled with tracer diffusion
measurements.

Finally, it would be interesting to vary the effective noise and diffusion
constants in a controlled manner in an experiment. While there is, as yet, no
method to calculate these constants directly from a microscopic theory it is
reasonable to expect that by decreasing the Peclet number (i.e., increasing the
role of {\em isotropic\/} thermal diffusion) one could drive the sedimenting
system into the unscreened phase.  Thus by repeating the experiments of
Segr\`{e} {\em et\ al.\/} \cite{Segre} with colloids that are more nearly
density matched to the solvent one could test our prediction of a transition to
an unscreened phase.

We would like thank M. Rutgers, P. Chaikin, and P. Segr\`{e} for communicating
unpublished results and for useful discussions.  We would also like to thank J.
Brady, D. Durian, E.  Herbolzheimer, S. Milner, R. Pandit, J. Rudnick and U.C.
T\"auber for useful discussions. S.R.  thanks F. Pincus and C. Safinya and the
Materials Research Laboratory, UCSB (NSF DMR93-01199 and 91-23045), as well as
the ITP Biomembranes Workshop (NSF PHY94-07194) for partial support in the
early stages of this work .  A.L.  acknowledges support by an AT\&T Graduate
Fellowship.  E.F. acknowledges support by a Heisenberg fellowship (FR 850/3-1)
from the Deutsche Forschungsgemeinschaft.


\begin{references}
\bibitem{freq} Including frequency-dependence will not alter the critical
  exponents and the structure of the scaling variable; it may, however, affect
  the functional from of the full relaxation rates.
\end{references}
\end{document}